\documentclass[%
prx,
 amsmath,amssymb,
 aps,
twocolumn,
longbibliography,
showkeys
]{revtex4-1}

\usepackage{graphicx}
\usepackage{epstopdf}
\usepackage{dcolumn}
\usepackage{bm}
\usepackage{color}
\usepackage[mathlines]{lineno}
\usepackage[breaklinks=true,colorlinks=true,linkcolor=blue,urlcolor=blue,citecolor=blue]{hyperref}
\usepackage{ragged2e}
\usepackage{enumitem}
\usepackage{bibunits}
\usepackage{amsthm}
\usepackage{diagbox}

\defaultbibliographystyle{elsarticle-num}
\defaultbibliography{biblio}

\newcommand{\lr}[1]{\left(#1\right)}

\renewcommand\thefigure{\arabic{figure}}

\begin{document}

\title[]{Stability of pairwise social dilemma games: destructive agents,  constructive agents, and their joint effects }

\author{Khadija Khatun$^{1,2}$}
\author{Chen Shen$^3$}
\email{steven\_shen91@hotmail.com}
\author{Lei Shi$^4$}
\email{shi\_lei65@hotmail.com}
\author{Jun Tanimoto$^{3,1}$}
\email{tanimoto@cm.kyushu-u.ac.jp}
\affiliation{
\vspace{2mm}
\mbox{1. Interdisciplinary Graduate School of Engineering Sciences, Kyushu University, Fukuoka, 816-8580, Japan}
\mbox{2. Faculty of Applied Mathematics Department, University of Dhaka, Dhaka-1000, Bangladesh}
\mbox{3. Faculty of Engineering Sciences, Kyushu University, Kasuga-koen, Kasuga-shi, Fukuoka 816-8580, Japan}
\mbox{4. School of Statistics and Mathematics, Yunnan University of Finance and Economics, Kunming, 650221, China}
}
\date{\today}

\begin{abstract}

Destructive agents, who opt out of the game and indiscriminately harm others, paradoxically foster cooperation, representing an intriguing variant of the voluntary participation strategy. Yet, their impact on cooperation remains inadequately understood, particularly in the context of pairwise social dilemma games and in comparison to their counterparts, constructive agents, who opt out of the game but indiscriminately benefit others. Furthermore, little is known about the combined effects of both agent types on cooperation dynamics. Using replicator dynamics in infinite and well-mixed populations, we find that, contrary to their role in facilitating cooperation in multiplayer games, destructive agents fail to encourage cooperation in pairwise social dilemmas. Instead, they destabilize and may even replace defection in the prisoners' dilemma and stag-hunt games. Similarly, in the chicken game, they can destabilize or replace the mixed equilibrium of cooperation and defection, and they undermine cooperation in the harmony game. Conversely, constructive agents, when their payoffs exceed their contributions to opponents, can exhibit effects similar to destructive agents. However, if their payoffs are lower, while they destabilize defection in prisoners' dilemma and stag-hunt games, they do not disrupt the cooperation equilibrium in harmony games and have a negligible impact on the coexistence of cooperation in chicken games. The combination of destructive and constructive agents does not facilitate cooperation but instead generates complex evolutionary dynamics, including bi-stable, tri-stable, and quad-stable states, with outcomes contingent on their relative payoffs and game types. These results, taken together, enhance our understanding of the impact of the voluntary participation mechanism on cooperation, contributing to a more comprehensive understanding of its influence.

\end{abstract}


\maketitle

\section*{Introduction}
The persistence of cooperative behavior poses a significant evolutionary puzzle. Cooperation often incurs costs for individuals to help others, while the temptation of free-riding—benefiting from others' assistance without contributing—threatens to undermine cooperative efforts~\cite{axelrod1981evolution,fehr2000cooperation}. According to the principle of `survival of the fittest', free riding, which saves the cost of helping, should have more evolutionary advantages than cooperation, leading to the latter's eventual extinction~\cite{darwin1964origin}. Evolutionary game theory offers a robust mathematical framework to unravel this paradox~\cite{weibull1997evolutionary,smith1982evolution}. In particular, a public goods game (PGG) is a mathematical metaphor for exploring the cooperation conundrum in multiplayer games~\cite{fischbacher2001people,archetti2012game}. In the PGG, cooperators invest in a common pool by incurring costs, whereas defectors contribute nothing. The cumulative payoff in the common pool is then multiplied by an enhancement factor and distributed to all participants, irrespective of their contribution. In scenarios where the game is one-shot and anonymous~\cite{fehr2002altruistic,fowler2010cooperative}, meaning that players never interact with the same individual more than once, and reciprocity mechanisms like reputation~\cite{ohtsuki2004should,ohtsuki2006leading}, costly signals~\cite{gintis2001costly,jordan2016third}, and repeated interactions~\cite{axelrod1980effective} are absent, fostering cooperation becomes particularly challenging~\cite{rand2013human}. In such contexts, social mechanisms such as reward~\cite{andreoni2003carrot,hilbe2010incentives,wang2018exploiting}, punishment~\cite{fehr2002altruistic,Dreber2008winners,li2018punishment,wang2017onymity}, social exclusion~\cite{sasaki2013evolution,li2024antisocial}, prior commitment\cite{han2017evolution,duffy2002actions}, and voluntary participation~\cite{hauert2002volunteering,szabo2002phase} become crucial for the emergence of cooperative behavior.

While social punishment (and reward) has fostered cooperation, its efficacy relies on identifying and tracking defectors. However, the stability of these mechanisms is threatened by second-order free-riders-those who contribute but avoid the costs of punishing (or rewarding)—and antisocial punishers (or rewarders)-those who defect yet punish (or reward) other defectors, potentially undermining the effectiveness of these social mechanisms~\cite{rand2010anti,rand2011evolution}. In contrast, voluntary participation emerges as a simple yet effective strategy that promotes cooperation without the complexities associated with identifying and tracking defectors~\cite{hauert2002volunteering,hauert2002replicator}. Importantly, this social mechanism does not face the same evolutionary challenges as punishment and reward, making it a subject of extensive study. Voluntary participants, also known as loners who abstain from partaking in the benefits generated from public goods and instead receive a fixed positive payoff by opting out, can effectively establish cooperation. This is achieved through a cyclic dominance effect, where cooperation yields to defectors, who, in turn, give way to loners, and loners give way to cooperators.  Extending beyond the original research, studies have explored the effects of loners in networked populations~\cite{hauert2005game,jia2023interactive}, the role of loners in punishment dilemmas~\cite{hauert2007via}, and various other cooperation-related issues~\cite{inglis2016presence}. Moreover, researchers have investigated different variants of the loner strategy, such as abstention strategies, where individuals neither pay nor receive anything while their opponents bear a participation cost~\cite{perez2022cooperation}. Exiters, who receive a fixed payoff but contribute nothing to their opponents, also receive attention~\cite{shen2021exit,shen2023exit}. Studies investigate the freedom to choose between homogeneous symmetric/asymmetric public resources~\cite{salahshour2021evolution,salahshour2021freedom}, hedgers who enact tit-for-tat play without cooperation in the first move~\cite{guo2020novel}, and other related aspects~\cite{wang2021super,guo2022effect}. Although these variants differ from the loner strategy, they all demonstrate a cooperation-promotion effect.

An intriguing variant of the loner strategy is represented by destructive agents, who, like loners, abstain from participating in public goods but actively harm others without personal gain. These agents can create stable cycles of cooperation, defection, and destruction in both finite and infinite populations, paradoxically promoting cooperation through their indiscriminate harmful actions~\cite{arenas2011joker,requejo2012stability}. This interesting result leads us to several intriguing questions. First, how do destructive agents impact cooperation dynamics in pairwise social dilemma games, where distinct equilibrium points exist (e.g., the dominance of cooperation in the harmony game, defection in the prisoner's dilemma, bistable equilibrium in the stag-hunt game, and mixed strategies equilibrium in the chicken game), compared to their effects in PGGs, which only exhibit cooperation and defection equilibria? Second, in contrast to destructive agents, what would be the impact of constructive agents, who positively contribute to both cooperators and defectors, on cooperation dynamics? Lastly, it is crucial to explore the joint effects of constructive and destructive agents on cooperation dynamics in social dilemma games, especially regarding how the presence of constructive agents may alter the influence of destructive agents on cooperation dynamics.

To explore these questions, we extend the framework of social dilemma games to incorporate both destructive and constructive agents. Initially, we analyze their effects on promoting cooperation within well-mixed populations separately, before investigating their combined impact. Our model incorporates key parameters like dilemma strength ($D_g$, $D_r$), categorizing games into harmony, chicken, stag-hunt, and prisoner's dilemma, along with incentives for agents to exit the game $d$, and the respective damage $d_1$ and benefit $d_2$ caused by destructive and constructive agents. Utilizing replicator dynamical equations, we discover that destructive agents fail to encourage cooperation in pairwise social dilemmas in contrast to their role in promoting cooperation in public goods games. Instead, they destabilize defection, ultimately replacing it in prisoner's dilemma and stag-hunt games while undermining cooperation in chicken and harmony games. Conversely,  constructive agents sustain the coexistence of cooperation in the chicken game and minimally influence the cooperative equilibrium in the harmony game, particularly when their payoffs are less than their contributions to opponents. Otherwise, their impact tends to mimic that of destructive agents. When both constructive and destructive agents are active simultaneously, their combined influence often mirrors the effects observed when each agent type acts alone. For example, the coexistence of destructive and constructive agents can disrupt defection in the prisoner's dilemma and stag-hunt games, while also compromising cooperation in chicken and harmony games. Furthermore, in scenarios where constructive agents confer benefits exceeding their gains, these joint effects can lead to the emergence of complex dynamics, including bi-stable, tri-stable, or quad-stable equilibria, contingent on game types and parameter conditions. These results enhance our understanding of the impact of the voluntary participation mechanism on cooperation, contributing to a more comprehensive understanding of its influence.

\section*{Model}\label{Model}

Our method contains two necessary basic components: (a) payoff matrices and (b) population settings and game dynamics. A brief description of each section is given as follows:
\subsection*{Payoff matrices}
In this study, we assume a symmetric pairwise game, where the evolutionary dynamics of cooperation within dyadic interactions involve the strategic interplay of cooperation ($C$) and defection($D$). In instances where both players opt for cooperation, they are endowed with the payoff denoted as $R$ (Reward). Conversely, if both players choose defection, the resulting payoff is designated as $P$ (Punishment). When one player cooperates while the other defects, two distinct payoffs emerge: $T$, representing the temptation to defect, signifying an advantageous outcome for the defector; and $S$, denoting the sucker's payoff, indicating a disadvantageous outcome for the cooperator. Based on the relative ordering of these payoffs, four types of social dilemma games can be identified: the prisoner's dilemma, characterized by $T>R>P>S$; the stag hunt, characterized by $R>T>P>S$; the chicken or snowdrift game, characterized by $T>R>S>P$; and the harmony game, characterized by $R>T>S>P$. 

To observe cooperation dynamics we have used the concept of universal scaling of dilemma strength ~\cite{wang2015universal}, where $D_g = T - R$ and $D_r = P - S$ are used to quantify the game's dilemma strength, encapsulating aspects characteristic of both chicken-type dilemmas (originating from greed) and stag-hunt-type dilemmas (originating from fear).  The nature of the equilibrium depends on the signs of $D_g$ and $D_r$: a prisoner's dilemma scenario, where both $D_g$ and $D_r$ are positive, leads to mutual defection as the equilibrium state. A positive $D_g$ combined with a negative $D_r$, resembling the chicken game, results in a mixed equilibrium of cooperation and defection. The stag-hunt game, indicated by a negative $D_g$ and a positive $D_r$, presents a bi-stable equilibrium, where both mutual cooperation and mutual defection are stable strategies. Finally, in the harmony game scenario, where both $D_g$ and $D_r$ are negative, cooperation emerges as the dominant equilibrium strategy. 
\paragraph{Pairwise game with destructive agents (DA)}

Incorporating destructive agents named Joker, which inflicts equal damage on both cooperators and defectors, without receiving any benefit, was initially introduced in a public good game (PGG)~\cite{arenas2011joker}. In this study, we introduce destructive agents into the pairwise game as a third strategy with no payoff. Then, we relax the strong assumption(Joker doesn't receive any benefit) with a positive payoff from destructive agents. The benefit received by destructive agents playing with others is $d\in[0,1)$ and the damage that imposes on its opponents is $d_1\in[0,1)$. The payoff matrix is given in table \ref{t01}.
\begin{table}[!t]
\caption{\label{t01} Payoff matrix of social dilemma game for destructive agents.}
\begin{ruledtabular}
\begin{tabular}{cccc}
~   & $C$ & $D$  & $DA$  \\
\hline
$C$ & $1$ & $-D_r$  &$-d_1$\\

$D$ & $1+D_g$ & 0  &$-d_1$\\

$DA$ & $d$ & $d$  &$d$\\
\end{tabular}
\end{ruledtabular}
\end{table}

\paragraph{Pairwise game with  constructive agents(CA).}

 Constructive agents in pairwise games strive to equal benefits between cooperators and defectors and also receive some benefits in participation. The aid, normal players receive from playing with constructive agents is $d_2\in[0,1)$ and the benefit received by the constructive agent is the same as the destructive agent did i.e. $d\in[0,1)$. The payoff matrix is given as table \ref{t02}.

\begin{table}[!t]
\caption{\label{t02} Payoff matrix of social dilemma game for  constructive agents.}
\begin{ruledtabular}
\begin{tabular}{cccc}
~   & $C$ & $D$  & $CA$  \\
\hline
$C$ & $1$ & $-D_r$  &$d_2$\\

$D$ & $1+D_g$ & 0  &$d_2$\\

$CA$ & $d$ & $d$  &$d$\\
\end{tabular}
\end{ruledtabular}
\end{table}
\paragraph{Pairwise game in mixed of destructive and  constructive agents.}
To comprehensively assess the impact of both constructive and destructive agents, we synthesized the strategies outlined in Tables \ref{t01} and \ref{t02} to create a new payoff matrix. This matrix incorporates four strategies: cooperation ($C$), defection ($D$),  constructive agents ($CA$), and destructive agents ($DA$). The detailed interactions and resultant payoffs are presented in table \ref{t03}.
\begin{table}[!t]
\caption{\label{t03} Payoff matrix of social dilemma game for the combined effect of destructive and  constructive agents.}
\begin{ruledtabular}
\begin{tabular}{ccccc}
~   & $C$ & $D$  & $DA$ &$CA$ \\
\hline
$C$ & $1$ & $-D_r$  &$-d_1$ & $d_2$\\
$D$ & $1+D_g$ & 0  &$-d_1$ & $d_2$\\
$DA$ & $d$ & $d$  &$d$ & $d$\\
$CA$ & $d$ & $d$  &$d$ & $d$\\
\end{tabular}
\end{ruledtabular}
\end{table}
\subsection*{Population setting and game dynamics}
We consider a well-mixed and infinite population model, wherein individuals engage in random pairwise interactions with each other.
\paragraph{Destructive agent's game dynamics:}
Let $x, y, z$ denote the fractions of cooperation, $C$, defection, $D$, and destructive agent, $DA$ in the population. Where $0 \leq x, y, z \leq 1$, and $x+y+z=1$. The expected payoff for each player is given as:
\begin{equation}
\begin{array}{l}
\Pi_{C}  = x-D_ry-d_1z,\\
\Pi_{D}  = (1+D_g)x -d_1z,\\
\Pi_{DA}  = d.\\\\
\end{array}.
\label{eq01}
\end{equation}
The replicator equations are:
\begin{equation}
\begin{array}{l}
\dot{x} = x\lr{\Pi_{C}-\overline{\Pi}_{DA}}, \\
\dot{y} = y\lr{\Pi_{D}-\overline{\Pi}_{DA}}, \\
\dot{z} = z\lr{\Pi_{DA}-\overline{\Pi}_{DA}}.
\end{array}
\label{eq02}
\end{equation} 
where, $\overline{\Pi}_{DA}=x\Pi_{C}+y\Pi_{D}+z\Pi_{DA}$.
\paragraph{ Constructive agent's game dynamics:}
 Let $w$ denote the fractions of the constructive agent, $CA$ in the population, then $0 \leq x, y, w \leq 1$, and $x+y+w=1$. The expected payoff for each player and the replicator dynamics is given as eq.\ref{eq03} and eq.\ref{eq04} respectively.
\begin{equation}
\begin{array}{l}
\Pi_{C}  = x-D_ry+d_2w,\\
\Pi_{D}  = (1+D_g)x +d_2w,\\
\Pi_{CA}  = d.\\\\
\end{array}.
\label{eq03}
\end{equation}

\begin{equation}
\begin{array}{l}
\dot{x} = x\lr{\Pi_{C}-\overline{\Pi}_{CA}}, \\
\dot{y} = y\lr{\Pi_{D}-\overline{\Pi}_{CA}}, \\
\dot{w} = w\lr{\Pi_{CA}-\overline{\Pi}_{CA}}.
\end{array}
\label{eq04}
\end{equation} 
where, $\overline{\Pi}_{CA}=x\Pi_{C}+y\Pi_{D}+w\Pi_{CA}$.
\paragraph{Game dynamics of the joint effects of DA and CA:}
When both destructive and constructive agents simultaneously interact with the cooperation and defection, then $0 \leq x, y, z,w \leq 1$, and $x+y+z+w=1$. The expected payoff for each player is given as:
\begin{equation}
\begin{array}{l}
\Pi_{C}  = x-yD_r-d_1z+d_2w,\\
\Pi_{D}  = (1+D_g)x -d_1z+d_2w,\\
\Pi_{DA}  = d,\\
\Pi_{CA}  = d\\\\
\end{array}.
\label{eq05}
\end{equation}
The replicator equations are:
\begin{equation}
\begin{array}{l}
\dot{x} = x\lr{\Pi_{C}-\overline{\Pi}}, \\
\dot{y} = y\lr{\Pi_{D}-\overline{\Pi}}, \\
\dot{z} = z\lr{\Pi_{DA}-\overline{\Pi}},\\
\dot{w} = w\lr{\Pi_{CA}-\overline{\Pi}}.
\end{array}
\label{eq06}
\end{equation}  
where, $\overline{\Pi}=x\Pi_{C}+y\Pi_{D}+z\Pi_{DA}+w\Pi_{CA}$.
Detailed explanations of the equilibria and their stability of all replicator dynamics have been given in the Appendix. 

\begin{figure*}[!t]
    \centering
\includegraphics[width=1.0\linewidth]{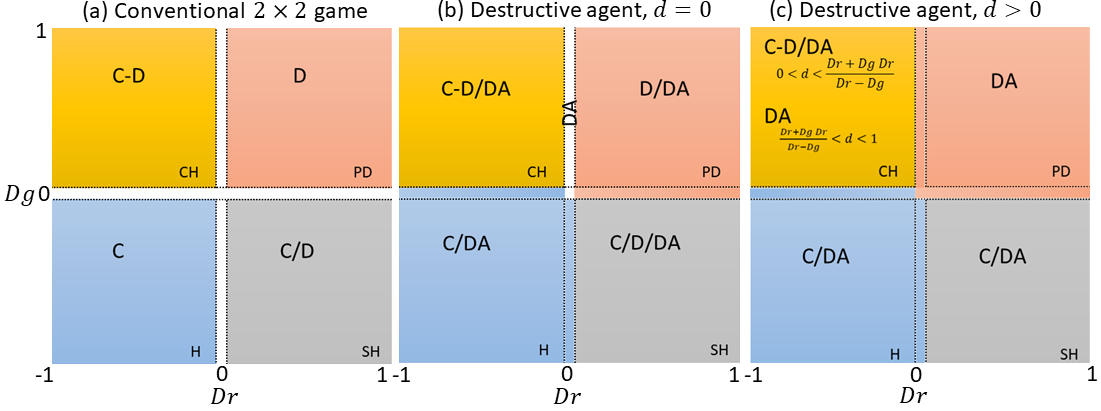}
    \caption{Non-beneficial destructive agents destabilize defection for $D_r>0$ , cooperation and a mix of cooperation and defection when $Dr<0$ (panel b). The defection of the prisoner’s dilemma is destabilized by a bi-stable defection and destruction, and Stag-Hunt's bi-stable equilibrium becomes tri-stable with destruction, on the other hand, Chicken's mixed cooperation and defection is transformed into a bi-stable mix of cooperation and defection or mono-morphic destruction, cooperation of Harmony turns into bi-stable cooperation and destruction. When the destructive agents receive a benefit (panel c), similar destabilization (or replacement when  $Dg>0$) is observed when $Dr<0$ but it replaces defection if $Dr>0$. Destruction replaces defection in the Prisoner's Dilemma and Stag-Hunt game and a mix of cooperation and defection in the Chicken game (after a threshold value of $d$). The diagrams can be divided into four regions(denoted by different colors) corresponding to Prisioner's Dilemma (PD), Stag hunt (SH), Harmony (H), and Chicken(CH) games, and the boundary ($D_r=0$ and $D_g=0$) separated by the black dotted lines. Equilibria, stable on the boundary are shown in the same color as the interior.    }
    \label{fig1}
\end{figure*}
\begin{figure*}[!t]
\includegraphics[width=1\linewidth]{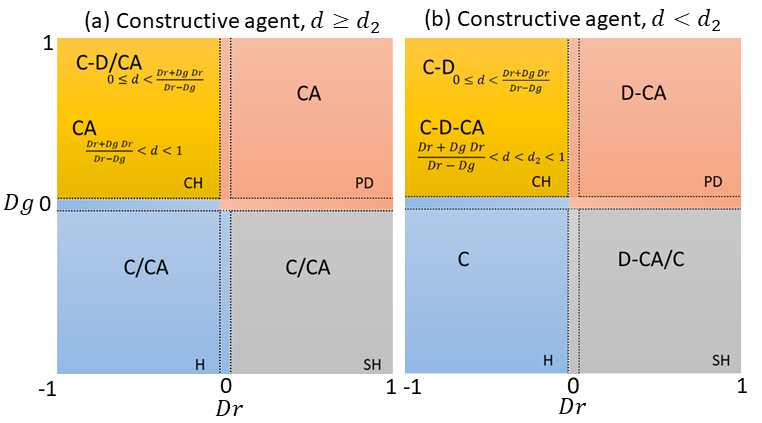}
    \caption{When the constructive agents' payoff exceeds others (panel a), construction replaces defection if $D_r>0$ and destabilizes cooperation and a mix of cooperation and defection if $D_r<0$. The stable equilibrium of Prisoner's Dilemma and Stag-Hunt is construction and a bi-stable of cooperation and construction. In contrast, Chicken's mixed equilibrium is bi-stable, either embracing a blend of cooperation and defection or construction or mono-stable construction (depending on $d$ values), and Harmony's cooperation demonstrates bi-stability with construction. When constructive agents' payoff is lower than others (panel b), defection is changed to polymorphic defection and construction if $Dr>0$ but does not influence cooperation and a mix of cooperation and defection if $Dr<0$. The defection of Prisoner's dilemma and  Stag-Hunt changes to a coexistence of defection and construction, and the stability of the equilibria remains unchanged in the Harmony and Chicken games.   }   
    \label{fig2}
\end{figure*}
\begin{figure*}[!t]
\includegraphics[width=1\linewidth]{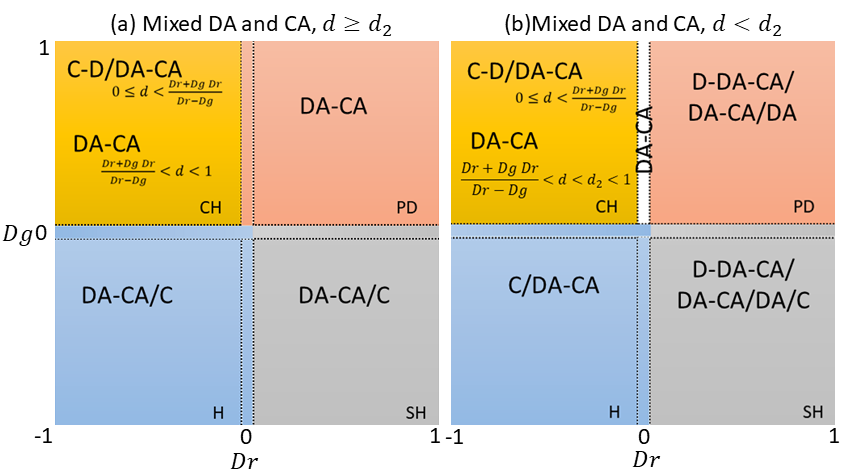}
    \caption{When the constructive agents' payoff exceeds others (panel (a)), a polymorphic mixture of destruction and construction replaces defection if $D_r>0$ and disrupts cooperation and a mix of cooperation and defection if $D_r<0$. The stable equilibrium of Prisoner's Dilemma and Stag-Hunt is a coexistence of destruction and construction and a bi-stable of cooperation and coexistence of destruction and construction. In contrast, Chicken's mixed equilibrium becomes either bi-stable, either embracing a blend of cooperation and defection or coexistence of destruction and construction or a mono-stable coexistence of destruction and construction, and Harmony's cooperation demonstrates bi-stability with the coexistence of destruction and construction. When constructive agents' payoff is lower than others (panel (b)), defection is changed to either coexistence of defection-destruction-construction or coexistence of destruction-cooperation or pure destruction if $D_r>0$ but if $D_r<0$ stability remains the same as panel (a).}   
    \label{fig3}
\end{figure*}
\section*{Results}\label{Results}
\subsection*{Destructive agents}
The presence of destructive agents in a PGG, paradoxically, promotes cooperation and destabilizes defection by cyclic dominance, where cooperation leads to defection, which leads to destruction, ultimately paving the way for cooperation again~\cite{arenas2011joker}. In contrast, the introduction of destructive agents in the prisoner's dilemma game—a special two-player (PGG)—fails to foster cooperation; instead, it destabilizes the equilibrium of the prisoner's dilemma game (refer to the upper right of Figure~\ref{fig1}b). In this scenario, the single defection equilibrium becomes bi-stable, with trajectories originating from an unstable node exhibiting two potential outcomes: direct destruction or defection overriding cooperation, as illustrated in the upper right of Figure~\ref{fig4}.

Beyond the prisoner's dilemma, our study extended to assess the influence of destructive agents within other pairwise social dilemma games, such as chicken, harmony, and stag-hunt. We analyzed their impact on game equilibria, focusing on mixed strategies of cooperation and defection, pure cooperation, and the bi-stable equilibrium between cooperation and defection. Our findings reveal that akin to observations in the prisoner's dilemma, destructive agents fail to promote cooperation; instead, they tend to destabilize existing equilibria (refer to Figure \ref{fig1}b for detailed illustrations). In the chicken game, the introduction of destructive agents transforms the mixed strategy equilibrium into a bi-stable system. This system is characterized by a possible coexistence of cooperation and defection, which is separated by a critical saddle point leading to destruction. The game dynamics evolve from two unstable equilibria towards these divergent outcomes, as depicted in the upper left panel of Figure~\ref{fig4}. The Harmony game's mono-stable cooperation becomes bi-stable with destructive agents, trajectories separated into either cooperation or destruction starting from two different unstable nodes(lower left of Figure~\ref{fig4}). The bi-stable cooperation or defection turns to a tri-stable by adding destructive agents in the stag-hunt game, trajectories from an unstable node are divided into three ways: cooperation, defection, and destruction (lower right panel of Figure~\ref{fig4}). 

The initial assumption regarding destructive agents posits that they receive no additional payoff from opting out, which can be seen as somewhat restrictive. Given the rarity of individuals who would opt out of the game without any potential benefits, we have decided to relax this assumption. Now, agents can derive benefits from opting out of the game. Similar to non-beneficial destructive agents, beneficial destructive agents do not facilitate cooperation. However, they can act as substitutes for defection in the prisoner's dilemma and stag-hunt games and destabilize equilibria in the harmony and chicken games, akin to the impact of non-beneficial destructive agents (as shown in Figure \ref{fig1}c). In the Prisoner's Dilemma game, defection is replaced by destruction; trajectories start from an unstable node directing to destruction directly or invading cooperation by defection, and defection by destruction (turning to the upper right of Figure~\ref{fig5}). In the Stag-Hunt game, the bi-stable cooperation or defection equilibrium shifts to the bi-stable equilibrium of cooperation or destruction; trajectories stemming from an unstable node present two possible outcomes: either direct cooperation or destruction, which prevails over defection. In the Chicken game, the mixed equilibrium is either similar to that of non-beneficial destructive agents (when $0\leq d <\frac{D_r+D_g D_r}{D_r-D_g}$) or mono-stable destruction ($ \frac{D_r+D_g D_r}{D_r-D_g}<d<1$; described in Appendix \ref{AppendixA}). Cooperation in the Harmony game produces the same outcome as the effect of non-beneficial destructive agents.

At a glance, destructive agents cannot promote cooperation in pairwise social dilemmas. However, they can destabilize and potentially replace defection in the prisoner's dilemma and stag hunt games; likewise, they can disrupt or supersede the mixed cooperation-defection equilibrium in the chicken game and undermine cooperation entirely in the harmony game. In contrast to destructive agents, which exploit or harm either cooperators or defectors, constructive agents emerge as a concept that benefits both parties equally and receives rewards for abstaining from participation. This introduces a new avenue of investigation into how constructive agents influence the dynamics of cooperation in pairwise social dilemma games, which we will explore further in subsequent analyses.

\subsection*{Constructive agents}
Similar to destructive agents, incorporating constructive agents in pairwise social dilemmas does not encourage cooperation. Rather, introducing these agents changes the stability of the equilibria in the dilemmas. Two distinct scenarios have been observed based on the relative payoffs received by constructive agents and the payoffs offered by constructive agents to others. When constructive agents experience greater payoff than the contributions they make to others, the destabilization and transformation of these agents mirror that of destructive agents, except that the outcome shifts from destruction to construction, as illustrated in the Figure~\ref{fig2}a and Figure~\ref{fig6} (theoretical analysis given in Appendix \ref{AppendixB}). 

Constructive agents, when receiving lower payoffs compared to the benefits they provide to others, disrupt defection equilibria in the prisoner's dilemma and stag hunt games. However, their introduction has no significant impact on cooperation in the harmony game and only a negligible effect on the coexistent equilibria of cooperation and defection in the Chicken game (see Figure~\ref{fig2}b). In the Prisoner's Dilemma game, when trajectories originate at an unstable equilibrium of purely constructive agents and sequentially lead to cooperation and then defection, the result is a polymorphic stable mix of defection and construction that supplants the mono-stable defection equilibrium (refer to the upper right of the Figure~\ref{fig7}). Similarly, in the stag-hunt game, the bi-stable equilibria of cooperation and defection become bi-stable cooperation or a polymorphic mixture of defection and construction(refer to the lower right of Figure~\ref{fig7}).  The mixed equilibria of chicken's analogously may be unchanged (when $0\leq d <\frac{D_r+D_g D_r}{D_r-D_g}$, see the analytical result in Appendix \ref{AppendixB} ) or shifted to polymorphic stable mixtures of cooperation, defection, and construction($ \frac{D_r+D_g D_r}{D_r-D_g}<d<1$, see upper left of the Figure~\ref{fig7}). 

To sum up, constructive agents, when their payoffs surpass their contributions to opponents, may demonstrate effects akin to destructive agents. Conversely, when their payoffs are lower, although they destabilize defection in prisoners' dilemma and stag-hunt games, they neither disturb cooperation in harmony games nor exert a significant influence on the coexistent equilibrium in chicken games. At this point, it is entirely natural to investigate the combined impact of both destructive and constructive agents.

\subsection*{Mixed of destructive agents and   constructive agents}
The introduction of both destructive and constructive agents in social dilemma games does not foster cooperation. Instead, it results in intricate evolutionary dynamics, where the end equilibrium is contingent on the relative payoff received by constructive agents and the payoffs offered by constructive agents to others. When the constructive agents' payoff exceeds the aids they have given to others, they displace defection fully in the Prisoners' Dilemma and  Stag-Hunt games and can destabilize cooperation and coexistent cooperation and defection in the Harmony and Chicken games(turn to Figure~\ref{fig3}a; see Appendix \ref{AppendixC} for theoretical analysis). In the Prisoner's Dilemma, defection is substituted by a coexistence of destruction and construction, turning to the upper right corner of Figure~\ref{fig8}. In this scenario, in simplex (C, DA, CA), for instance, all trajectories either converge to cooperation or coexistence of destruction and cooperation, an introduction of mutant defection can invade cooperation (refer to the simplex (C, D, DA) in the same figure), but not the mixture which leads the mixture as final equilibrium.  The bi-stable equilibrium of Stag-Hunt becomes bi-stable between cooperation and coexistence of destruction and construction (turn to the lower right of Figure~\ref{fig8}).  All trajectories divided by a collection of unstable nodes (simplex (C, DA, CA), for example, in the same figure), converge either towards cooperation or the coexistence of destruction and construction; the introduction of mutant defection is unable to infiltrate the stability, consequently, bi-stability between cooperation and the mix of destruction and construction sustained. Similarly, Chicken's mixed equilibrium may become bi-stable, encompassing either a mixture of cooperation and defection or destruction and construction (when $0\leq d <\frac{D_r+D_g D_r}{D_r-D_g}$; see upper left of Figure.~\ref{fig8}) or mono-stable a mixture of destruction and construction ($ \frac{D_r+D_g D_r}{D_r-D_g}<d<1$; see Appendix \ref{AppendixC}), and Harmony's cooperation exhibits bi-stability with a mix of destruction and construction(see lower left of Figure~\ref{fig8}).

However, when constructive agents receive lower payoffs than the benefits given to opponents, the equilibria in Prisoner's Dilemma and Stag Hunt shift to complex coexistence of defection, destruction, and construction, showing expanded multi-stability, while the equilibria in Harmony and Chicken remain unchanged as constructive agents have higher payoffs, illustrated in Figure~\ref{fig3}b and theoretical analysis in Appendix \ref{AppendixC}. In the Prisoner's Dilemma, the mono-stable defection equilibrium is replaced by either the coexistence of defection-destruction-construction or the coexistence of destruction-cooperation or pure destruction, exhibited in the upper right corner of Figure~\ref{fig9}. In this context, trajectories in simplex (C, DA, CA) are divided by a branch of unstable nodes into cooperation or a mix of destruction and cooperation, an introduction of mutant defection can invade cooperation to mix of defection, destruction, and construction (refer to the simplex (D, DA, CA)) or destruction only (in the simplex (C, D, DA)) in the same figure, but no influence on the mixture of destruction and cooperation, which leads a tri-stable state either coexistent of defection, destruction, and construction or a mix of destruction and construction or destruction only. Similarly, in the Stag-Hunt game, the bi-stable equilibria of cooperation and defection become tetra-stable cooperation or the coexistence of defection-destruction-construction or the coexistence of destruction-cooperation or pure destruction (refer to the lower right corner of Figure \ref{fig9}). In this scenario, trajectories within the simplex (C, DA, CA) are partitioned by a branch of unstable nodes, creating a bi-stability between cooperation and a combination of destruction and construction. The introduction of mutant defection does not invade cooperation but results in a bi-stable state, either a mixture of defection, destruction, and construction (observed in the simplex (D, DA, CA)) or destruction only (within the simplex (C, D, DA)) in the same figure. This mutant defection has no impact on the blend of destruction and construction, maintaining a quad-stable state that encompasses cooperation, the coexistence of defection, destruction, and construction, or a combination of destruction and construction, or destruction alone.  

\section*{Discussion}\label{Discussion}
To discuss, in this paper, we have demonstrated that, contrary to their role in facilitating cooperation within public goods games, the introduction of destructive agents into pairwise social dilemma games fails to encourage cooperation. Specifically, destructive agents, when deriving no benefit, destabilize the system. This leads to a shift from equilibria of single defection, single cooperation, or mixed states to regions of bi-stability or even tri-stability that include defection, cooperation, and destruction. In the prisoner's dilemma, harmony, and chicken games, we observe transitions to bi-stability involving defection and destruction, cooperation and destruction, and mixed states with destruction. In the stag-hunt game, a unique shift to tri-stability incorporating defection, cooperation, and destruction occurs. Conversely, when destructive agents gain benefits, they entirely displace the defection equilibrium in both prisoner's dilemma and stag-hunt games. 

Additionally, we introduced a novel agent type akin to destructive agents:  constructive agents. These agents exit the game upon receiving a benefit, yet they also endow their opponents with additional benefits. Our findings suggest that when constructive agents secure higher payoffs than those they bestow on opponents, they can destabilize defection in the prisoner's dilemma and stag-hunt games and disrupt cooperation in the chicken and harmony games, mirroring the destabilizing influence of destructive agents. However, if the payoff for constructive agents is less than what they provide to their opponents, they predominantly disrupt defection states. This leads to new equilibria where defection coexists with constructive actions in the prisoner's dilemma, and a bi-stable state between mixed defection and construction, and cooperation in the stag-hunt games, leaving the dynamics in the chicken and harmony games unaffected.

Moreover, combining destructive and constructive agents does not inherently promote cooperation but introduces more complex dynamics, especially when the payoff for constructive agents is lower than what they bestow upon opponents. For instance, in the prisoner's dilemma, a tri-stable state emerges, characterized by mixed defection, destructive, and constructive agents; a mixed state of destructive and constructive agents; and a state dominated by destructive agents. In the stag-hunt game, a quad-stable state arises, featuring mixed states of defection, destruction, and constructive agents; a mixed destructive and constructive agent state; a purely destructive state; and a state of pure cooperation. The harmony game exhibits bi-stability between pure cooperation and a mixed destructive and constructive agent state. In the chicken game, dynamics are parameter-dependent, sometimes resulting in bi-stability involving a mixed cooperation and defection state, and a mixed destructive and constructive agents state, or leading to a singular mixed state of destructive and constructive agents under different conditions.

The concept of loner strategy, alongside destructive and constructive agents, parallels the notion of social value orientation~\cite{murphy2011measuring}. In this framework, loners embody individualistic values, seeking personal payoff without impacting their opponents. Destructive agents align with competitive values, aiming to harm their opponents while securing non-negative benefits. Conversely,  constructive agents represent prosocial values by benefiting their opponents while also obtaining non-negative payoffs. While the influence of these strategies on cooperation has been extensively studied, the role of voluntary participation in fostering cooperation remains underexplored. These strategies, being specific, do not encapsulate the broader spectrum of potential behaviors. Beyond these, the social value orientation framework suggests additional motivations for innovative variants of voluntary strategies. These include masochism, where individuals accept negative payoffs by exiting the game without affecting others; martyrdom, which entails negative personal payoffs alongside generating positive outcomes for others; sadomasochism, characterized by negative personal payoffs coupled with inflicting harm on opponents; among others. Therefore, developing a comprehensive theoretical model that integrates a general voluntary participation strategy, rooted in social value orientations, presents a compelling research direction. This approach aims to investigate how diverse social values impact the evolution of cooperation and assess their effectiveness in enhancing cooperative behaviors. Such an endeavor is poised to deepen our understanding of how various voluntary participation strategies can address the enduring puzzle of cooperation.

The critical assumptions of this study—namely, one-shot, anonymous, and well-mixed scenarios—present a most challenging context for the evolution of cooperation. While we found that both constructive and destructive agents do not facilitate cooperation in the context of pairwise social dilemma games, the investigation of the impact of these agents warrants further exploration, as realistic situations often involve repeated interactions or some prior information. It is of significant interest to investigate the impact of these agents on cooperation dynamics in scenarios involving repeated interactions~\cite{rossetti2023direct}, networked populations~\cite{perc2017statistical,wang2015evolutionary}, higher-order interactions~\cite{guo2021evolutionary}, and other scenarios~\cite{XIA20238}.

\section*{Article information}
\paragraph*{Data Accessibility.} The programs for theoretical analysis and image generation are given at \url{https://osf.io/4p3ch}.

\paragraph*{Acknowledgements.} This research was supported by the National Natural Science Foundation of China (grant no.~11931015). We also acknowledge support from (i) a JSPS Postdoctoral Fellowship Program for Foreign Researchers (grant no. P21374), and an accompanying Grant-in-Aid for Scientific Research from JSPS KAKENHI (grant no. JP 22F31374) to C.\,S., (ii) the National Natural Science Foundation of China (grants no.~11931015,~12271471 and 11671348), and the major Program of National Fund of Philosophy and Social Science of China (grants no.~22\&ZD158 and 22VRCO49)  to L.\,S., and (iv) the grant-in-Aid for Scientific Research from JSPS, Japan, KAKENHI (grant No. JP 20H02314 and JP 23H03499) awarded to J.\,T, (v) Japanese Government (MEXT)
scholarship, Japan, ( grant No. 222143) awarded to K.\,K.
\paragraph*{Author contributions.} 
C.\,S. and J.\,T. conceived research. K.\,K. and C.\, S. performed analytical analysis. All co-authors discussed the results and wrote the manuscript.
\paragraph*{Conflict of interest.} Authors declare no conflict of interest.

\onecolumngrid
\renewcommand\theequation{A\arabic{equation}}
\setcounter{equation}{0}
\setcounter{figure}{0}
\renewcommand\thefigure{A\arabic{figure}}
\section*{Appendix}

\subsection{Equilibria and Stability of destructive agent}\label{AppendixA}
 Four realistic equilibrium points exist in the presence of destructive agents obtained from the solution of the replicator dynamics eq.~\ref{eq02}: $E_{A1}=(1,0,0)$, $E_{A2}=(0,1,0)$, $E_{A3}=(0,0,1)$, $E_{A4}=(\frac{-D_r}{D_g-D_r},\frac{D_g}{D_g-D_r},0)$. \\
 
Firstly we reduce the system of equations into a lower dimension, setting $z=1-x-y$ into eq.~\ref{eq02}  then the new set of equations will be: 
\begin{equation}
\begin{array}{l}
\dot{x} = x\lr{(1-x)\lr{\Pi_{C}-\Pi_{DA}}-y\lr{\Pi_{D}-\Pi_{DA}}}=f_{C}(x,y), \\
\dot{y} = y\lr{(1-y)\lr{\Pi_{D}-\Pi_{DA}}-x\lr{\Pi_{C}-\Pi_{DA}}}=f_{D}(x,y). \\

\end{array}
\label{eq07}
\end{equation}  
To examine the stability of these equilibrium points, we calculate the eigenvalues of the Jacobin matrix:
\begin{equation}
J_{A}=
\begin{bmatrix}
\frac{\partial f_{C}(x,y)}{\partial x} & \frac{\partial f_{C}(x,y)}{\partial y}\\
\frac{\partial f_{D}(x,y)}{\partial x}&\frac{\partial f_{D}(x,y,)}{\partial y}
\end{bmatrix}
\label{eq08}
\end{equation}
Where,
\begin{equation}
\begin{cases}
\begin{aligned}
\frac{\partial f_{C}(x,y)}{\partial x} = &- \left((-d + (1 + D_g) x - d_1 (1 - x - y)) y) + (1 - x) (-d + x- 
    d_1 (1 - x-y)-D_ry)\right.\\  
& \left. +x(d+(1+d)(1-x)-x+d_1(1-x-y)-(1+ d_1 +D_g)y+D_ry)\right.,\\
\frac{\partial f_{C}(x,y)}{\partial y}= & x\left (d + (d_1 - D_r) (1-x)-(1+D_g) x+d_1(1-x-y)-d_1 y)\right), \\
\frac{\partial f_{D}(x,y)}{\partial x}= & y\left (d -x-(1+d_1)x +(1+d_1+D_g) (1-y)+d_1(1-x-y)+D_r )\right), \\
\frac{\partial f_{D}(x,y)}{\partial y}= & \left(-d + (1 + D_g) x - d_1 (1 - x - y)) (1-
    y) +(d -(1 + D_g) x -(d- D_r)x\right.\\ 
& \left. +d_1 (1 - y) + d_1 (1 - x - y)) y - x (-d + x - d_1 (1-x-y)-D_r y)\right).
\end{aligned}
\end{cases}
\label{eq09}
\end{equation}
 
For a dynamical system represented by its equilibrium points, stability ~\cite{anagnost1991stability} analysis involves examining the real parts of its eigenvalues. If all eigenvalues possess negative real parts, the equilibrium is deemed stable due to the system's tendency to return to this state over time. Conversely, if any eigenvalue has a positive real part, the equilibrium becomes unstable, indicating divergence from the steady state. When eigenvalues include negative real parts and those with real parts equal to zero, necessitating a deeper analysis, applying the center manifold theorem~\cite{carr2012applications} becomes crucial to understanding the system's behavior near that particular point.
\subparagraph{Stability of the equilibria:}
\begin{enumerate}
\item  $E_{A1}$: $\lambda_1=D_g$ and $\lambda_2=-1+d$, so the real parts of the eigenvalues will be negative if $d<1$ and $D_g<0$. hence, the equilibrium point $E_{A1}$ is stable if $D_g<0$. However, at $D_g=0$, we find a zero eigenvalue, to conclude the stability of this point we need to use the center manifold theorem here.  
    The Jacobin matrix at $E_{A1}$ is:
    \begin{equation}
J_{A1}=
\begin{bmatrix}
-1+d & -1+d\\
0&0
\end{bmatrix}
\label{eq10}
\end{equation}
An invertible matrix $U$ is constructed by arranging the eigenvectors of the matrix \( J_{A1} \) as its column elements, which can diagonalize the matrix
\begin{equation}
U=
\begin{bmatrix}
1 & 0\\
-1&1
\end{bmatrix}
\label{eq011}
\end{equation}
Therefor,\begin{equation}
U^{-1} J_{A1} U=
\begin{bmatrix}
-1+d & 0\\
0&0
\end{bmatrix}
\label{eq012}
\end{equation} 

The new coordinates are eq.~\ref{eq013}  and the eq.~\ref{eq07} has been transformed into eq.~\ref{eq14} 
\begin{equation}
\begin{bmatrix}
u\\
v
\end{bmatrix}
=U^{-1}\begin{bmatrix}
x\\
y
\end{bmatrix} =
\begin{bmatrix}
x+y\\
y
\end{bmatrix}
\label{eq013}
\end{equation}
\begin{equation}
\begin{array}{l}
\dot{u} = (-1 + u) (d u - d_1 (-1 + u) u - (u - v) (u - D_r v)), \\
\dot{v} = -v (d + d_1 (-1 + u)^2 - d u + (u - v) (-1 + u - D_r v)). \\
\end{array}
\label{eq14}
\end{equation}  
Set $u=u_1+1$, then the eq.~\ref{eq14} is converted to a diagonal form eq.~\ref{eq15} 
\begin{equation}
\begin{array}{l}
\dot{u_1} = u_1 (-((1 + d_1) u_1^2) + d (1 + u_1) - (-1 + v) (-1 + D_r v) + 
   u_1 (-2 - d_1 + v + D_r v)), \\
\dot{v} = -v ((1 + d_1) u_1^2 + D_r (-1 + v) v - u_1 (-1 + d + v + D_r v)). \\
\end{array}
\label{eq15}
\end{equation}
which can be written as:
\begin{equation}
\begin{array}{l}
\dot{X} = P X+F(X,Y) \\
\dot{Y} = Q Y+G(X,Y) \\
\end{array}
\label{eq16}
\end{equation}
Here, $X=v$, $Y=u_1$, and  $P=0$, $Q=-1+d$; $F$ and $G$ are functions of $X$ and $Y$ and $ F(0) = G(0) = 0, F'(0) = G'(0) = 0$, there exists a $\delta>0$ 
 and a function $h\in C^r(N_\delta(0) ), \forall  r\geq1$,  so that $h(0)=h'(0)=0$ defines the local center manifold $\{(X,Y)\in R^2|u_1=h(v) for |v|<\delta\}$  and satisfies \\
 $h'(v)[ P v+F(v,h(v))]=Q h(v)+G(v,h(v))$.\\
Set $u_1=O(v^2)$, then we obtain 
\begin{equation}
\begin{array}{l}
\dot{v} = D_r v^2+O(v^3).
\end{array}
\label{eq17}
\end{equation}
If $D_r<0$, the central manifold will be stable at the origin. So we can say that at $D_g \leq 0$, $E_{A1}$ will be stable when $D_r<0$.\\
   
\item $E_{A2}$: $\lambda_1=d$ and $\lambda_2=-D_r$, unstable for all $d>0$.  If $d=0$, $E_{A2}$ has a zero eigenvalue with a negative eigenvalue for $D_r>0$. In a similar process, we find the following transformed system eq.~\ref{eq18} and the center manifold eq.~\ref{eq19}
\begin{equation}
\begin{array}{l}
\dot{u} = u (D_g u (1 + u + v) - D_r (1 + u) (1 + u + v) + v (u - d_1 v)), \\
\dot{v} = v ((D_g - D_r) u^2 + (1 + D_g - D_r) u (1 + v) - d_1 v (1 + v)). \\
\end{array}
\label{eq18}
\end{equation}
\begin{equation}
\begin{array}{l}
\dot{v} = -d_1 v^2+O(v^3).
\end{array}
\label{eq19}
\end{equation}
Which is stable at the origin, so the equilibrium point $E_{A2}$ is stable when $d=0$ and $D_r>0$.\\
 \item $E_{A3}$: $\lambda_1=-d-d_1$ and $\lambda_2=-d-d_1$, stable for all $d>0$
 \item $E_{A4}$: $\lambda_1=\frac{D_r+d*D_g-d*D_r+D_g*D_r}{D_g-D_r}$ and $\lambda_2=\frac{D_g*D_r}{D_g-D_r}$,  will be stable if $D_r<0$, $D_g>0$ and $0\leq d\leq\frac{-(D_r + D_g D_r)}{D_g-D_r}$.
\end{enumerate}
\subsection{Equilibria and Stability of  constructive agent}\label{AppendixB}
 There are six realistic equilibrium points in the presence o  constructive agents obtained from the solution of replicator dynamics eq.~\ref{eq04}: $E_{B1}=(1,0,0)$, $E_{B2}=(0,1,0)$, $E_{B3}=(0,0,1)$, $E_{B4}=(\frac{-D_r}{D_g-D_r},\frac{D_g}{D_g-D_r},0)$, $E_{B5}=(0,\frac{d_2-d}{d_2},\frac{d}{d_2})_{d_2>d}$ , and $E_{B6}=(\frac{D_r(d-d_2)}{d_2D_g + D_r - d_2 D_r + D_g D_r},\frac{-D_g(d-d_2)}{d_2 D_g + D_r - d_2 D_r + D_g D_r},\frac{d D_g + D_r - d D_r + D_g D_r}{d_2 D_g + D_r - d_2 D_r + D_g D_r})_{d_2>d}$. \\
 
Similarly, we reduce the system of equations into a lower dimension, setting $w=1-x-y$ into eq.~\ref{eq04} then the new set of equations will be: 
\begin{equation}
\begin{array}{l}
\dot{x} = x\lr{(1-x)\lr{\Pi_{C}-\Pi_{A}}-y\lr{\Pi_{D}-\Pi_{CA}}}=g_{C}(x,y), \\
\dot{y} = y\lr{(1-y)\lr{\Pi_{D}-\Pi_{A}}-x\lr{\Pi_{C}-\Pi_{CA}}}=g_{D}(x,y). \\

\end{array}
\label{eq20}
\end{equation}  

To examine the stability of these equilibrium points, we calculate the eigenvalues of the Jacobin matrix:
\begin{equation}
J_{B}=
\begin{bmatrix}
\frac{\partial g_{C}(x,y)}{\partial x} & \frac{\partial g_{C}(x,y)}{\partial y}\\
\frac{\partial g_{D}(x,y)}{\partial x}&\frac{\partial g_{D}(x,y,)}{\partial y}
\end{bmatrix}
\label{eq21}
\end{equation}
Where,
\begin{equation}
\begin{cases}
\begin{aligned}
\frac{\partial g_{C}(x,y)}{\partial x} = &- \left((-d + (1 + D_g) x+d_2(1-x-y)) y)+(1-x)(-d+x+d_2(1-x-y)-D_r y)\right.\\  
& \left. +x(d+(1-d_2)(1-x)-x-d_2(1-x-y)-(1-d_2+D_g)y+D_r y)\right.,\\
\frac{\partial g_{C}(x,y)}{\partial y}= & x\left(d +(-d_2-D_r) (1-x)-(1+D_g)x-d_2 (1-x-y)+d_2y)\right), \\
\frac{\partial g_{D}(x,y)}{\partial x}= & y\left (d-x-(1-d_2)x+(1-d_2+D_g)(1-y) -d_2(1-x-y)+D_r y)\right), \\
\frac{\partial g_{D}(x,y)}{\partial y}= & \left(-d+(1+D_g)x+d_2(1-x-y)) (1-y)+(d-(1+D_g)x-(-d_2-D_r)x \right.\\ 
& \left. -d_2(1-y)-d_2 (1-x-y)) y-x (-d+x+d_2 (1-x-y)-D_r y)\right).
\end{aligned}
\end{cases}
\label{eq22}
\end{equation}

\subparagraph{Stability of the equilibria:}
\begin{enumerate}
\item  $E_{B1}$: $\lambda_1=D_g$ and $\lambda_2=-1+d$, so the real parts of the eigenvalues will be negative if $d<1$ and $D_g<0$. hence, the equilibrium point $E_{A1}$ is stable if $D_g<0$. However, at $D_g=0$, we find a zero eigenvalue, to conclude the stability of this point we need to use the center manifold theorem. We can find the transformed system in eq.~\ref{eq22} and the center manifold eq.~\ref{eq23} in the previous way.\\
\begin{equation}
\begin{array}{l}
\dot{u} = u ((-1 + d2)u^2 +d(1 + u)-(-1 + v)(-1 + D_r v)+u (-2+d_2+v+D_rv)), \\
\dot{v} = v ((-1+d_2) u^2-D_r (-1+v) v+u(-1+d+v+D_r v)). \\
\end{array}
\label{eq22}
\end{equation}
\begin{equation}
\begin{array}{l}
\dot{v} = D_r v^2+O(v^3).
\end{array}
\label{eq23}
\end{equation}\\
The coefficient of $v^2$ will be negative if $D_r<0$, and the center manifold is stable at the origin. Hence the point $E_{B1}$ is stable when $D_g\leq 0$ and $D_r<0$.\\
\item $E_{B2}$: $\lambda_1=d$ and $\lambda_2=-D_r$, unstable for all $d>0$.  If $d=0$, then $E_{B2}=(0,1,0)$ has a zero eigenvalue with a negative eigenvalue for $D_r>0$. In a similar process, we find the following transformed system eq.~\ref{eq24} and the center manifold eq.~\ref{eq25}
\begin{equation}
\begin{array}{l}
\dot{u} = u (D_g u (1 + u + v) - D_r (1 + u) (1 + u + v) + v (u + d_2 v)), \\
\dot{v} = v ((D_g - D_r) u^2 + (1 + D_g - D_r) u (1 + v) + d_2 v (1 + v)). \\
\end{array}
\label{eq24}
\end{equation}
\begin{equation}
\begin{array}{l}
\dot{v} = d_2 v^2+O(v^3).
\end{array}
\label{eq25}
\end{equation}
Which is unstable at the origin as the coefficient of $v^2$ is positive for $0\leq d_2 <1$, so the equilibrium point $E_{B2}$ is unstable when $d  \geq 0$ and $D_r>0$.\\
 \item $E_{B3}$: $\lambda_1=-d+d_2$ and $\lambda_2=-d+d_2$, is stable for all $-1\leq D_g, D_r\leq1$ if $d_2<d$ and unstable otherwise.
 \item $E_{B4}$: $\lambda_1=\frac{D_r+d*D_g-d*D_r+D_g*D_r}{D_g-D_r}$ and $\lambda_2=\frac{D_g*D_r}{D_g-D_r}$,  will be stable if $D_r<0$, $D_g>0$ and $d\leq\frac{-D_r(1+D_g)}{D_g-D_r}$.
 \item $E_{B5}$: $\lambda_1=\frac{-d(d_2-d)}{d_2}$ and $\lambda_2=\frac{-D_r(d_2-d)}{d_2}$, will be stable if
 $-1\leq D_g\leq 1$, $0 < D_r \leq 1$ and $0\leq d <d_2$.
  \item $E_{B6}$: $\lambda_1=\frac{(-d+d_2) D_g D_r (d_2 D_g +D_r-d_2 D_r+D_g D_r)}{(-d_2 D_g - D_r + d_2 D_r - D_g D_r)^2}$ and $\lambda_2=-\frac{(-d+d_2) (d D_g+D_r-d D_r+D_g D_r) (d_2 D_g +D_r-d_2 D_r+ D_g D_r)}{(-d_2 D_g - D_r + d_2 D_r - D_g D_r)^2}$, will be stable if $-1 < D_r < 0 $, $0 < D_g \leq 1$, and $\frac{D_r + D_g D_r}{-D_g + D_r} < d < d_2<1$.
\end{enumerate}
\subsection{Equilibria and Stability of the joint of destructive  and constructive agent}\label{AppendixC}
In combination with destructive agents and constructive agents, there are seven realistic equilibrium points obtained from the solution of the replicator dynamics eq.~\ref{eq06}:
$E_{C1}=\lr{0, 0, a, 1-a}_{a\in[0, 1]}$,  $E_{C2}=\lr{a, 0, \frac{-d+d_2+a-d_2 a}{d_1 + d_2}, \frac{d+d_1-a-d_1 a}{d_1 + d_2}}_{a\in(0, 1)}$, $E_{C3}=\lr{0, 1, 0, 0}$, $E_{C4}=\lr{1, 0, 0, 0}$, $E_{C5}=\lr{\frac{-D_r}{D_g-D_r}, \frac{D_g}{D_g-D_r}, 0, 0}$, $E_{C6}=\lr{0,a,\frac{-d+ d_2 -d_2 a}{d_1 + d_2},\frac{d+d_1-d_1 a}{d_1 + d_2}}_{a\in(0, 1)}$,  and $E_{C7}=\lr{a,\frac{-D_g a}{D_r}, \frac{-d D_r + d_2 D_r+d_2 D_g a+D_r a-d_2 D_r a+D_g D_r a}{(d_1 + d_2) D_r}, \frac{d D_r+d_1 D_r+d_1 D_g a-D_r a-d_1 D_r a-D_g D_r a}{(d_1 + d_2) D_r}}_{a\in(0, 1)}$.\\

Set $w=1-x-y-z$ into the eq.~\ref{eq06}, then the system will be:
\begin{equation}
\begin{array}{l}
\dot{x} = x\lr{(1-x)\lr{\Pi_{C}-\Pi_{CA}}-y\lr{\Pi_{D}-\Pi_{CA}}-z\lr{\Pi_{DA}-\Pi_{CA}}}=h_{C}(x,y,z), \\
\dot{y} = y\lr{(1-y)\lr{\Pi_{D}-\Pi_{CA}}-x\lr{\Pi_{C}-\Pi_{CA}}}=h_{D}(x,y,z), \\
\dot{z} = z\lr{-y\lr{\Pi_{D}-\Pi_{CA}}-x\lr{\Pi_{C}-\Pi_{CA}}}=h_{J}(x,y,z). \\
\end{array}
\label{eq26}
\end{equation}
The Jacobin matrix is:
\begin{equation}
J_{C}=
\begin{bmatrix}
\frac{\partial h_{C}(x,y,z)}{\partial x} & \frac{\partial h_{C}(x,y,z)}{\partial y}&\frac{\partial h_{C}(x,y,z)}{\partial z}\\
\frac{\partial h_{D}(x,y,z)}{\partial x}&\frac{\partial h_{D}(x,y,z)}{\partial y}&\frac{\partial h_{D}(x,y,z)}{\partial z}\\
\frac{\partial h_{J}(x,y,z)}{\partial x}&\frac{\partial h_{J}(x,y,z)}{\partial y}&\frac{\partial h_{J}(x,y,z)}{\partial z}
\end{bmatrix}
\label{eq27}
\end{equation}
 Where, 
 \begin{equation}
\begin{cases}
\begin{aligned}
\frac{\partial h_{C}(x,y,z)}{\partial x} = &- \left. y(-d+(1+D_g) x+d_2 (1-x-y-z)-d_1 z)+(1-x) (-d+x-
D_r y+d_2 (1-x-y-z)-d_1z) \right.\\  
& \left.+x (d+(1-d_2) (1-x)-x-(1-d_2+D_g) y+D_r y-   d_2 (1-x-y-z)+d_1z)\right.,\\
\frac{\partial h_{C}(x,y,z)}{\partial y}= & x\left. (d+(-d_2-D_r) (1-x)-(1+D_g) x+d_2 y-d_2 (1-x-y-z)+   d_1 z)\right.), \\
\frac{\partial h_{C}(x,y,z)}{\partial z}= & x\left ((-d_1-d_2) (1-x)-(-d_1-d_2) y)\right), \\
\frac{\partial h_{D}(x,y,z)}{\partial x}= & y \left.(d-x- (1-d_2)x+(1-d_2+D_g) (1-y) + D_r y - 
   d_2 (1-x-y-z)+d_1 z) \right. ,\\ 
\frac{\partial h_{D}(x,y,z)}{\partial y}=& \left. (1-y) (-d+(1+D_g) x+d_2 (1-x-y-z)-d_1 z)-x (-d+x-D_r y +d_2 (1-x-y-z)-d_1 z)\right),\\
& \left. +y (d-(1+D_g) x-(-d_2-D_r) x-d-2 (1-y)-d_2 (1-x-y-z)+ d_1 z) \right.,\\
\frac{\partial h_{D}(x,y,z)}{\partial z}=& \left. (-(-d_1-d_2) x)+(-d_1-d_2) (1-y)) y\right.,\\
\frac{\partial h_{J}(x,y,z)}{\partial x}=& \left.z (d-x-(1-d_2) x-(1-d_2+D_g) y+D_r y-d_2 (1-x-y-z)+d_1 z)\right.,\\
\frac{\partial h_{J}(x,y,z)}{\partial y}=&\left. z (d-(1+D_g) x-(-d_2-D_r) x+d_2 y-d_2 (1-x-y-z)+d_1 z)\right.,\\
\frac{\partial h_{J}(x,y,z)}{\partial z}=&\left. (-((-d_1-d_2) x)-(-d_1-d_2) y) z-y (-d+(1 + D_g)x+d_2(1-x-y-z)-d_1 z)-x (-d+x-D_r y \right.\\
& \left. + d_2 (1-x-y-z)-d_1 z) \right. \\
\end{aligned}
\end{cases}
\label{eq28}
\end{equation}
\subparagraph{Stability of the equilibria:}
\begin{enumerate}
    \item At $E_{C1}$: $\lambda_{1,2}=-d+d_2-a (d_1+d_2)$ and $\lambda_3=0$, are the eigenvalues, the real parts of $\lambda_{1,2}<0$ for $0\leq d_1<1$, if $0\leq d_2 \leq d < 1$ and $ a>0$ or if $0 \leq d < d_2$ and $a > \frac{(-d +d_2}{d_1+d_2})$. Since there is a zero eigenvalue to conclude we have to use the center manifold theorem here.\\
    The Jacobin matrix at $E_{C1}=\lr{0, 0, a, 1-a}$ is:
    \begin{equation}
J_{C1}=
\begin{bmatrix}
-d-a d_1 +(1-a) d_2, & 0 &0\\
0&-d-a d_1 +(1-a) d_2 & 0\\
a (d+a d_1-(1-a) d_2) & a (d+a d_1-(1-a) d_2) &0
\end{bmatrix}
\label{eq29}
\end{equation}
An invertible matrix $U$ is constructed by arranging the eigenvectors of the matrix \( J_{C1} \) as its column elements, which can diagonalize the matrix
\begin{equation}
U=
\begin{bmatrix}
-\frac{1}{a} & -1 & 0\\
0&1&0\\
1&0&1
\end{bmatrix}
\label{eq30}
\end{equation}
Therefor,\begin{equation}
U^{-1} J_{C1} U=
\begin{bmatrix}
-d+d_2-a (d_1+d_2) & 0 &0\\
0&-d+d_2-a (d_1+d_2)&0\\
0&0&0
\end{bmatrix}
\label{eq31}
\end{equation} 

The new coordinates are eq.~\ref{eq32}  and the eq.~\ref{eq26} has been transformed into eq.~\ref{eq33} 
\begin{equation}
\begin{bmatrix}
u\\
v\\
w1
\end{bmatrix}
=U^{-1}\begin{bmatrix}
x\\
y\\
z
\end{bmatrix} =
\begin{bmatrix}
-a (x+y)\\
y\\
a(x+y)+z
\end{bmatrix}
\label{eq32}
\end{equation}

\begin{equation}
\begin{aligned}
\dot{u} = &\left. \frac{1}{a^2}(a+ u)\lr{ (-1 + d_2) u^2 + a^2 (D_g - D_r) v^2 - 
   a u \lr{d + d_1 u + v - D_g v + D_r v + d_1 w_1 + d_2 (-1 + u + w_1)}} ,\right. \\
\dot{v} = & \left. \frac{1}{a^2}v ((-1 + d_2) u^2 - 
   a u (1 + d + D_g + d_1 u + v - D_g v + D_r v + d_1 w_1 + 
      d_2 (-2 + u + w_1)\right.\\
      & \left. -a^2 (d + d_1 u + v + D_g v - D_g v^2 + D_r v^2 + d_1 w_1 + 
    d_2 (-1 + u + w_1))),\right. \\
\dot{w_1} =  & \left.-\frac{1}{a^2}(a - w_1) \lr{(-1 + d_2) u^2 + a^2 (D_g - D_r) v^2 - 
   a u (d + d_1 u + v - D_g v + D_r v + d_1 w_1 + d_2 (-1 + u + w_1)}.\right. \\
\end{aligned}
\label{eq33}
\end{equation}
Set $w_1=w+a$, then the system \ref{eq33} in $ (u,v,w)$ will be:
\begin{equation}
\begin{aligned}
\dot{u} = &\left. -\frac{1}{a^2}(a+ u)\lr{ -((-1 + d_2) u^2) + a^2 (d_1 u + d_2 u + (-D_g + D_r) v^2) + 
 a u (d + d_1 u + v - D_g v + D_r v + d_1 w + d_2 (-1 + u + w))} ,\right. \\
\dot{v} = & \left. \frac{1}{a^2}v ((-1 + d_2) u^2 - 
 a u (1 + d + D_g + d_1 u + v - D_g v + D_r v + d_1 (a + w) + 
    d_2 (-2 + a + u + w))\right.\\
      & \left. -a^2 (d + d_1 u + v + D_g v - D_g v^2 + D_r v^2 + d_1 (a + w) + 
   d_2 (-1 + a + u + w))),\right. \\
\dot{w} =  & \left.-\frac{1}{a^2} w \lr{-((-1 + d_2) u^2) + a^2 (d_1 u + d_2 u + (-D_g + D_r) v^2) + 
   a u (d + d_1 u + v - D_g v + D_r v + d_1 w + d_2 (-1 + u + w))}.\right. \\
\end{aligned}
\label{eq34}
\end{equation}
The eq.~\ref{eq34} can be written as:
\begin{equation}
\begin{array}{l}
\dot{X} = P X+F(X,Y) \\
\dot{Y} = Q Y+G(X,Y) \\
\end{array}
\label{eq35}
\end{equation}
Here, $X=w$, $Y=
\begin{bmatrix}
u\\
v
\end{bmatrix}
$, and  $P=0$, $Q=
\begin{bmatrix}
-d+d_2-a (d_1+d_2) & 0 \\
0&-d+d_2-a (d_1+d_2)
\end{bmatrix}
$; $F$ and $G$ are functions of $X$ and $Y$and $ F(0) = G(0) = 0, F'(0) = G'(0) = 0$, there exists a $\delta>0$ and a function $H\in C^r(N_\delta(0) ), \forall  r\geq1$,  so that $H(0)=H'(0)=0$ defines the local center manifold $\{(X,H(X))\in R^3|Y=H(w) for |w|<\delta\}$  and satisfies \\
 $H'(w)[ P w+F(w,H(w))]=Q H(w)+G(w,H(w))$.\\ 
 
 Set $Y=O(w^2)$, we find the following center manifold: 
 \begin{equation}
\begin{array}{l}
\dot{w} = - \frac{1}{a} \lr{a (d_1+d_2)+(d-d_2)} w^3+O(w^4).
\end{array}
\label{eq36}
\end{equation}\\

The coefficient of $w^3$ will be negative for either $d>d_2$ or $d<d_2$ for all $ 0\leq d, d_1, d_2 <1$, so the center manifold is stable at the origin. Hence, the equilibrium point $E_{C1}$ is stable for all $ 0\leq d, d_1, d_2 <1$.
     \item At $E_{C2}$: eigenvalues are $\lambda_1=0$, and $\lambda_2=a(1-d)$, $\lambda_3= a D_g$. Here $\lambda>0$ for all $0 \leq d<1$, so the equilibrium point is unstable. 
 \item $E_{C3}$: eigenvalues are $\lambda_{1, 2}=d$ and $\lambda_3=-D_r$, so $E_{C3}$ is unstable as eigenvalues are positive for $d>0$.
 \item $E_{C4}$: $\lambda_{1, 2}=-1+d$ and $\lambda_3=D_g$, the real part of the eigenvalues will be negative if $d<1$ and $D_g<0$, if $D_g=0$ then one eigenvalue will be zero. Using the center manifold theorem, we obtain the following transformed system eq.~\ref{eq37} and the center manifold is eq.~\ref{eq38}.
 \begin{equation}
\begin{array}{l}
\dot{u}=(1 + u) (d v + d_1 (1 + u) v - v^2 + d_2 v (u + v) + v w + D_r v w - 
   D_r w^2),\\
\dot{v}=(-1 + v) (d v + d_2 u v + d_1 (1 + u) v - v^2 + d_2 v^2 + v w + D_r v w - 
   D_r w^2),\\
\dot{w} = -w (d + d_1 (1 + u) - v - d v - d_1 (1 + u) v + v^2 - 
   d_2 (-1 + v) (u + v) + w - v w - D_r v w + D_r w^2).
\end{array}
\label{eq37}
\end{equation}
\begin{equation}
\begin{array}{l}
\dot{w} = -(d+d_1) w+O(w^2).
\end{array}
\label{eq38}
\end{equation}\\
The center manifold is stable at the origin, which implies $E_{C4}$ is stable when $D_g\leq 0$ and $-1 \leq D_r \leq 1$ for all $0 \leq d, d_1, d_2 <1$. 
 \item $E_{C5}$: $\lambda_{1, 2}=\frac{D_r+d*D_g-d*D_r+D_g*D_r}{D_g-D_r}$ and $\lambda_3=\frac{D_g*D_r}{D_g-D_r}$, the real parts of the eigenvalues will be negative if $D_r<0$, $D_g>0$ and $d\leq\frac{-D_r(1+D_g)}{D_g-D_r}$, so $E_{C5}$ will be stable.
  \item $E_{C6}$:  $\lambda_1=-d+d_2$, $\lambda_2=- a d$, and $\lambda_3=-a D_r$ are the eigenvalues, the real parts of $\lambda_2 <0$ and $\lambda_3<0$ if $D_r>0$. To conclude the stability rather analytic way we rely on the numerical procedure to avoid complexity (see Figure.~\ref{fig9}). It is stable if $0<d<d_2$, $D_r>0$ and $-1\leq D_g\leq 1$.\\
  \item $E_{C7}$: Coexistent of all strategies, we also rely on numerical process to conclude this point's stability. It's unstable for all possible values of the parameters. 
  
\end{enumerate}
\clearpage

\begin{figure}[!t]
    \centering
\includegraphics[width=0.86\linewidth]{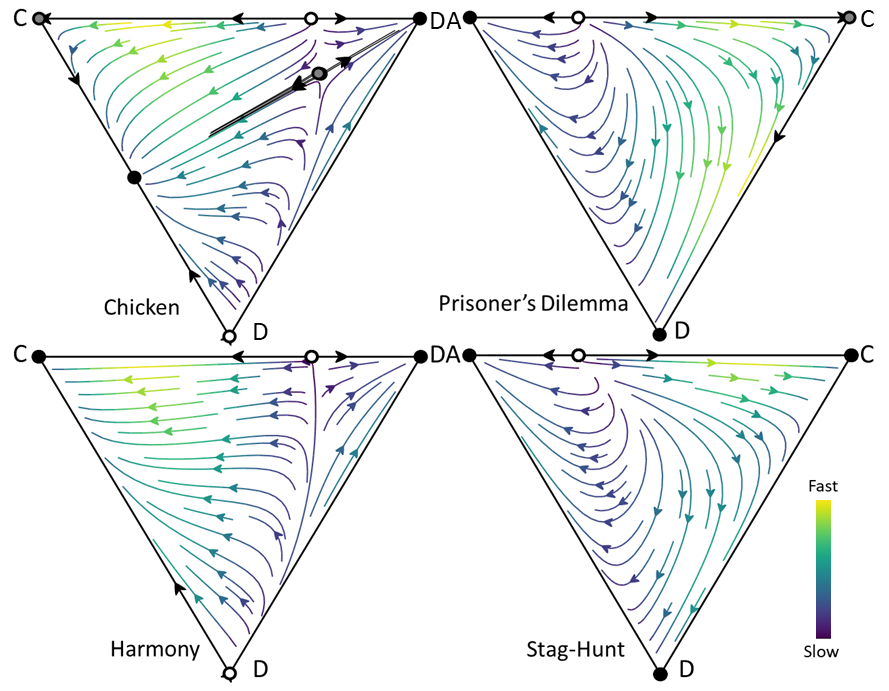}
    \caption{The stable defection of the Prisoner’s dilemma is replaced by a bi-stable defection and destruction, the Chicken's mixed equilibrium of cooperation and defection is transformed into a bi-stable either a mix of cooperation and defection or mono-morphic destruction, cooperation of Harmony turns into bi-stable cooperation and destruction, and Stag-Hunt bi-stable equilibrium becomes tri-stable with destruction. The parameters are fixed at $D_g=D_r=0.5, d_1=0.4$, and $d=0.0$. Solid black dots are stable nods, whites are unstable nods and grays are saddle points. Images are generated by a modified version of the 'egttools' Python Package~\cite{FERNANDEZDOMINGOS2023106419}. }
    \label{fig4}
\end{figure}

\begin{figure}[htbp]
    \centering
\includegraphics[width=1.0\linewidth]{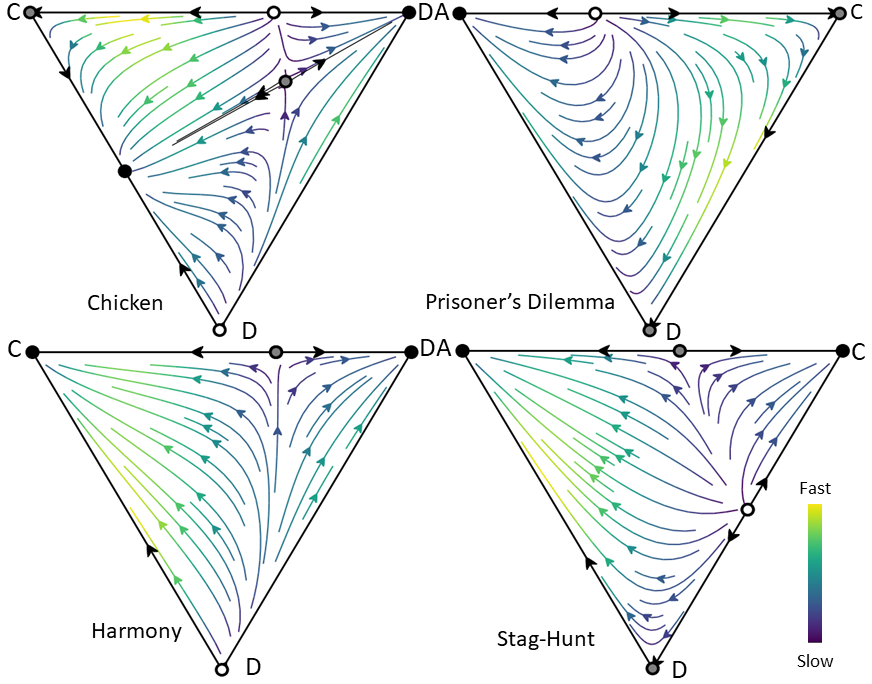}
    \caption{The Prisoner’s Dilemma game's stable equilibrium is destruction rather than defection, Chicken's mixed equilibrium changes to either a bi-stable mixer of cooperation and defection, and destruction or mono-stable destruction, and cooperation of Harmony turns into bi-stable cooperation and destruction, and Stag-Hunt's bi-stable equilibrium of cooperation and defection becomes bi-stable cooperation and destruction. The parameters are fixed at $D_g=D_r=0.5, d_1=0.4$, and $d=0.1$. Solid black dots are stable nods, whites are unstable nods and grays are saddle points.
    }
    \label{fig5}
\end{figure}
\begin{figure}[htbp]
    \centering
\includegraphics[width=0.89\linewidth]{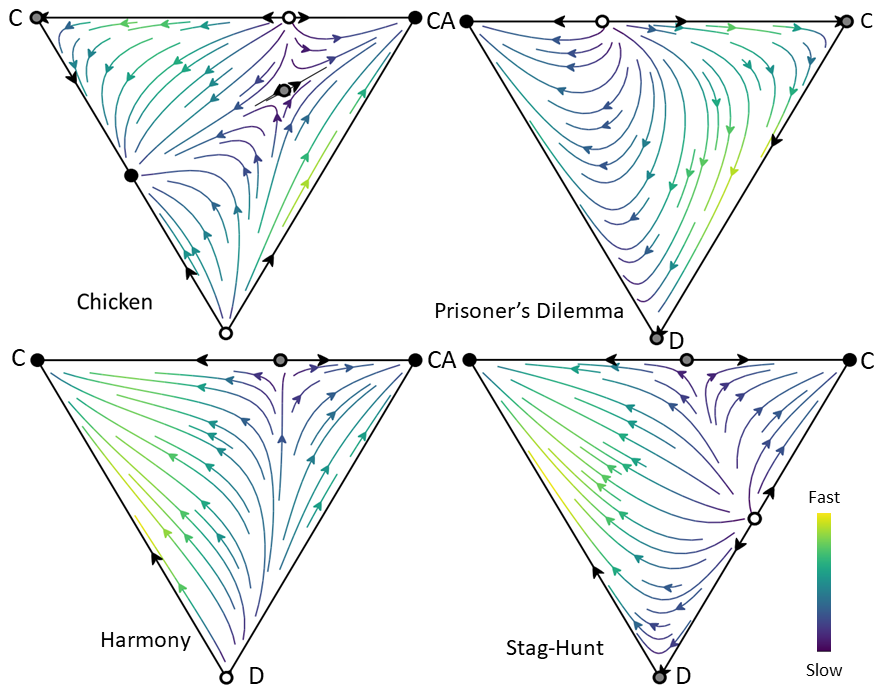}
    \caption{When constructive agents achieve higher payoffs than others, both defection and cooperation are destabilized by it. In the Prisoner's Dilemma and Stag-Hunt stability of defection is replaced with construction, while  Chicken's mixed equilibrium becomes bi-stable, either embracing a blend of cooperation and defection or construction, and Harmony's cooperation demonstrates bi-stability with construction.  The parameters are fixed at $D_g=D_r=0.5, d=0.4$, and $d_2=0.1$.Stable nodes are marked with solid black dots, unstable nodes with white dots, and saddle points with gray dots. }
    \label{fig6}
\end{figure}
\begin{figure}[!t]
    \centering
\includegraphics[width=0.86\linewidth]{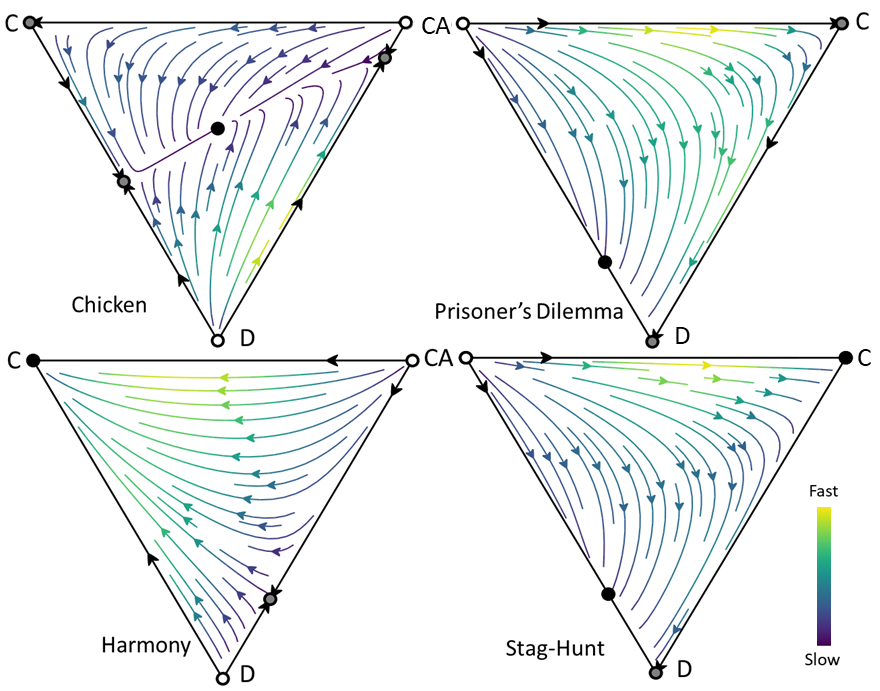}
    \caption{ Coexistence of construction with cooperation and defection in Chicken game and disruption of defection states by a mixture of defection and construction in Prisoner's Dilemma and Stag-Hunt, no influence in Harmony's cooperation. The parameters are fixed at $D_g=D_r=0.5, d=0.1$, and $d_2=0.4$.  Stable nodes are marked with solid black dots, unstable nodes with white dots, and saddle points with gray dots. }
    \label{fig7}
\end{figure}
\begin{figure*}[htbp]
    \centering
\includegraphics[width=0.89\linewidth]{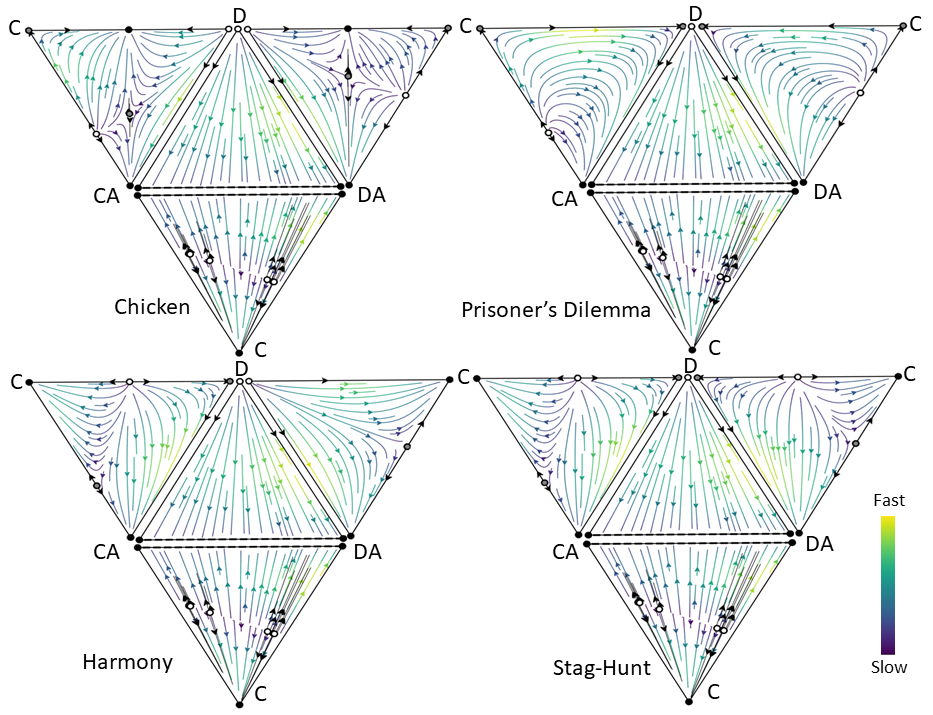}
    \caption{The mixture of destruction and construction shifts defection in the Prisoner's Dilemma and Stag-Hunt and destabilizes cooperation and coexistent cooperation and defection in Harmony and Chicken game. Four three-simplex combined as a four-simplex; for instance, in Prisoner's Dilemma simplex (C, DA, CA) is bi-stable cooperation and mix of destruction and construction, a mutant defection can invade cooperation and leads to a mono-stable mixture of destruction and destruction.   The parameters are fixed at $D_g=D_r=0.5, d=0.4$, $d_2=0.1$, and $d_2=0.1$.  Stable nodes are marked with solid black dots, all points are stable in the thick black dashed line,  unstable nodes with white dots, and saddle points with gray dots.}
    \label{fig8}
\end{figure*}
\begin{figure*}[htbp]
    \centering
\includegraphics[width=0.89\linewidth]{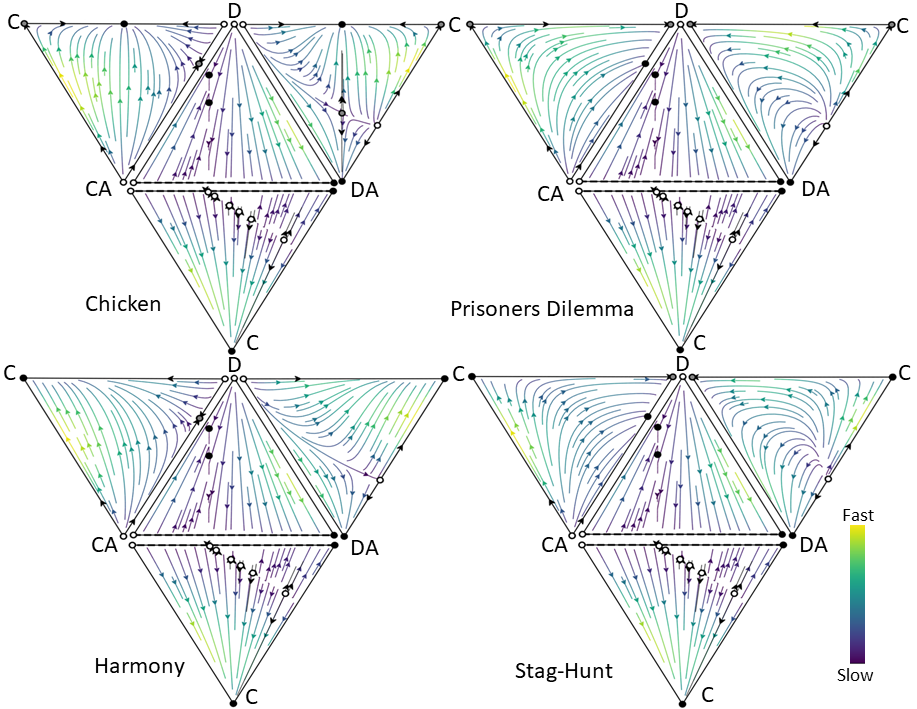}
    \caption{In the Prisoner's Dilemma, the mono-stable defection equilibrium is replaced by either the coexistence of defection-destruction-construction or the coexistence of destruction-cooperation or pure destruction, and the Stag-Hunt game, the bi-stable equilibria of cooperation and defection become tetra-stable cooperation or the coexistence of defection-destruction-construction or the coexistence of destruction-cooperation or pure destruction. Destabilization of cooperation and coexistent cooperation and defection also take place in the Harmony and Chicken game as in the previous one. In the Prisoners Dilemma, simplex (D, DA, CA) is a tri-stable coexistence of defection, destruction, and construction, the coexistence of destruction and construction, and destruction, a mutant cooperation cannot change the stability as it is invaded by defection.  The parameters are fixed at $D_g=D_r=0.5, d=0.1$, $d_1=0.4$, and $d_2=0.4$.  Stable nodes are marked with solid black dots, all points are stable in the thick black dashed line,  unstable nodes with white dots, and saddle points with gray dots.}
    \label{fig9}
\end{figure*}
\clearpage
\bibliography{biblio}
\bibliographystyle{elsarticle-num}

\end{document}